\documentclass[twocolumn,amsmath,amssymb,prl,superscriptaddress,aps]{revtex4-1}
 \pdfoutput=1 
\usepackage{longtable}
\usepackage{bm}
\usepackage{dcolumn}
\usepackage{epsfig}
\usepackage{graphicx}
\usepackage{subfigure}
\usepackage{comment}
\usepackage{epstopdf}
\usepackage{amsfonts}
\usepackage{amsmath}
\usepackage{amssymb}
\usepackage{subfigure}
\usepackage{color}

\usepackage[colorlinks,bookmarks=false,citecolor=blue,linkcolor=blue,urlcolor=blue]{hyperref}

\newcommand{\be}{\begin{equation}}
\newcommand{\ee}{\end{equation}}
\newcommand{\bea}{\begin{eqnarray}}
\newcommand{\eea}{\end{eqnarray}}

\usepackage{times}
\usepackage{appendix}

\begin{document}
 \pdfoutput=1 
%

\title{    Covalency and vibronic couplings    make a nonmagnetic $j\!=\!3/2$ ion magnetic}

\author{Lei Xu}
\email{Correspondence:l.xu@ifw-dresden.de }
\affiliation {Institute for Theoretical Solid State Physics, IFW Dresden, Helmholtzstr.~20, 01069 Dresden, Germany}

\author{Nikolay A.~Bogdanov}
\affiliation {Institute for Theoretical Solid State Physics, IFW Dresden, Helmholtzstr.~20, 01069 Dresden, Germany}

\author{Andrew Princep}
\affiliation {Department of Physics, University of Oxford, Clarendon Laboratory, Parks Road, Oxford, OX1 3PU, United Kingdom}

\author{Peter Fulde}
\affiliation{Max-Planck-Institut f\"{u}r Physik komplexer Systeme, N\"{o}thnitzer Str.~38, 01187 Dresden, Germany}

\author{Jeroen van den Brink}
\affiliation {Institute for Theoretical Solid State Physics, IFW Dresden, Helmholtzstr.~20, 01069 Dresden, Germany}
\affiliation {Institute for Theoretical Physics, TU Dresden, 01069 Dresden, Germany}

\author{Liviu Hozoi}
\email{Correspondence: l.hozoi@ifw-dresden.de}
\affiliation {Institute for Theoretical Solid State Physics, IFW Dresden, Helmholtzstr.~20, 01069 Dresden, Germany}

\begin{abstract}
\noindent
{\bf {\large ABSTRACT}}  \\

For 4$d^1$ and 5$d^1$ spin-orbit-coupled electron configurations, the notion of nonmagnetic $j\!=\!3/2$
quartet ground state discussed in classical textbooks is at odds with the observed variety of
magnetic properties.
Here we throw fresh light on the electronic structure of 4$d^1$ and 5$d^1$ ions in molybdenum- and
osmium-based double-perovskite systems and reveal different kinds of on-site many-body physics in the
two families of compounds:
while the sizable magnetic moments and $g$ factors measured experimentally are due to both
metal\,$d$\,--\,ligand\,$p$ hybridization and dynamic Jahn-Teller interactions for 4$d$ electrons,
it is essentially $d$\,--\,$p$ covalency for the 5$d^1$ configuration.
These results highlight the subtle interplay of spin-orbit interactions, covalency and electron-lattice
couplings as the major factor in deciding the nature of the magnetic ground states of 4$d$ and 5$d$
quantum materials.
Cation charge imbalance in the double-perovskite structure is further shown to allow a fine
tuning of the gap between the $t_{2g}$ and $e_g$ levels, an effect of much potential in the context
of orbital engineering in oxide electronics.
\end{abstract}

\date\today
\maketitle

\noindent
{\bf {\large INTRODUCTION}}  \\

A defining feature of $d$-electron systems is the presence of sizable electron correlations,
also referred to as Mott-Hubbard physics.
The latter has been traditionally associated with first-series (3$d$) transition-metal (TM)
oxides.
But recently one more ingredient entered the TM-oxide `Mottness' paradigm --- large spin-orbit
couplings (SOC's) in 4$d$ and 5$d$ quantum materials \cite{IrO_mott_kim_08,IrO_balents_review_14}.
It turns out that for specific $t_{2g}$-shell electron configurations, a strong SOC can
effectively augment the effect of Hubbard correlations \cite{IrO_mott_kim_08}:
although the 4$d$ and 5$d$ orbitals are relatively extended objects and the Coulomb repulsive
interactions are weakened as compared to the more compact 3$d$ states, the spin-orbit-induced
level splittings can become large enough to break apart the `nonrelativistic' $t_{2g}$ bands into
sets of well separated, significantly narrower subbands for which even a modest Hubbard $U$
acting on the respective Wannier orbitals can then open up a finite Mott-Hubbard-like gap
\cite{IrO_mott_kim_08}.
On top of that, SOC additionally reshuffles the intersite superexchange \cite{Ir213_KH_jackeli_09}.
The surprisingly large anisotropic magnetic interactions that come into play via the strong 
SOC's in iridates \cite{Ir213_KH_jackeli_09,Ir213_KH_gretarsson_2013,Ir213_KH_BJKim_2015,Ir213_yamaji_2014,Na2IrO3_vmk_14,Ir214_bogdanov_15},
for example, are responsible for an exotic assortment of novel magnetic ground states and 
excitations \cite{Ir213_KH_jackeli_09,Ir213_KH_chaloupka_10,IrO_balents_review_14}.

For large $t_{2g}$--$e_g$ splittings, the spin-orbit-coupled $t_{2g}^1$ and $t_{2g}^5$ electron
configurations can be in first approximation viewed as `complementary': in the simplest picture,
the $d$-shell manifold can be shrunk to the set of $j\!=\!1/2$ and $j\!=\!3/2$ relativistic levels,
with a $j\!=\!3/2$ ground state for the TM $t_{2g}^1$ configuration and a $j\!=\!1/2$ ground
state for $t_{2g}^5$ \cite{d1_SOC_kotani_1949,book_abragam_bleaney,book_mabbs_machin}.
While strongly spin-orbit-coupled $t_{2g}^5$ oxides and halides --- iridates, rhodates and
ruthenates, in particular --- have generated substantial experimental and theoretical
investigations in recent years, much of the properties of 5$d$ and 4$d$ $t_{2g}^1$ systems
remain to large extent unexplored.

From textbook arguments \cite{d1_SOC_kotani_1949,book_abragam_bleaney,book_mabbs_machin}, the 
$t_{2g}^1$ $j\!=\!3/2$ quadruplet should be characterized by a vanishing magnetic moment in cubic
symmetry, due to perfect cancellation of the spin and angular momentum contributions.
But this assertion leaves unexplained the wide variety of magnetic properties recently found in
4$d^1$ and 5$d^1$ cubic oxide compounds
\cite{BaYMoO_cussen_2006,BaYMoO_aharen_2010,BaYMoO_deVries_2010,BaYMoO_carlo_2011,BaYMoO_vib_qu_2013,Stitzer_DPs,BaMOsO_erickson_2007,BaMOsO_steele_2011}.
Ba$_2$YMoO$_6$, for example, develops no magnetic order despite a Curie-Weiss temperature of
$\approx$--200\,K \cite{BaYMoO_cussen_2006,BaYMoO_aharen_2010} and features complex magnetic
dynamics that persists down to the mK range, possibly due to either a valence-bond-glass 
\cite{BaYMoO_deVries_2010} or spin-liquid \cite{BaYMoO_carlo_2011} ground state.
Also Ba$_2$NaOsO$_6$ displays an antiferromagnetic Curie-Weiss temperature \cite{Stitzer_DPs,BaMOsO_erickson_2007}
but orders ferromagnetically below 7\,K \cite{BaMOsO_steele_2011} whilst Ba$_2$LiOsO$_6$ is a
spin-flop antiferromagnet \cite{BaMOsO_steele_2011}.

\begin{figure}[b]
\includegraphics[scale=0.35,angle=0]{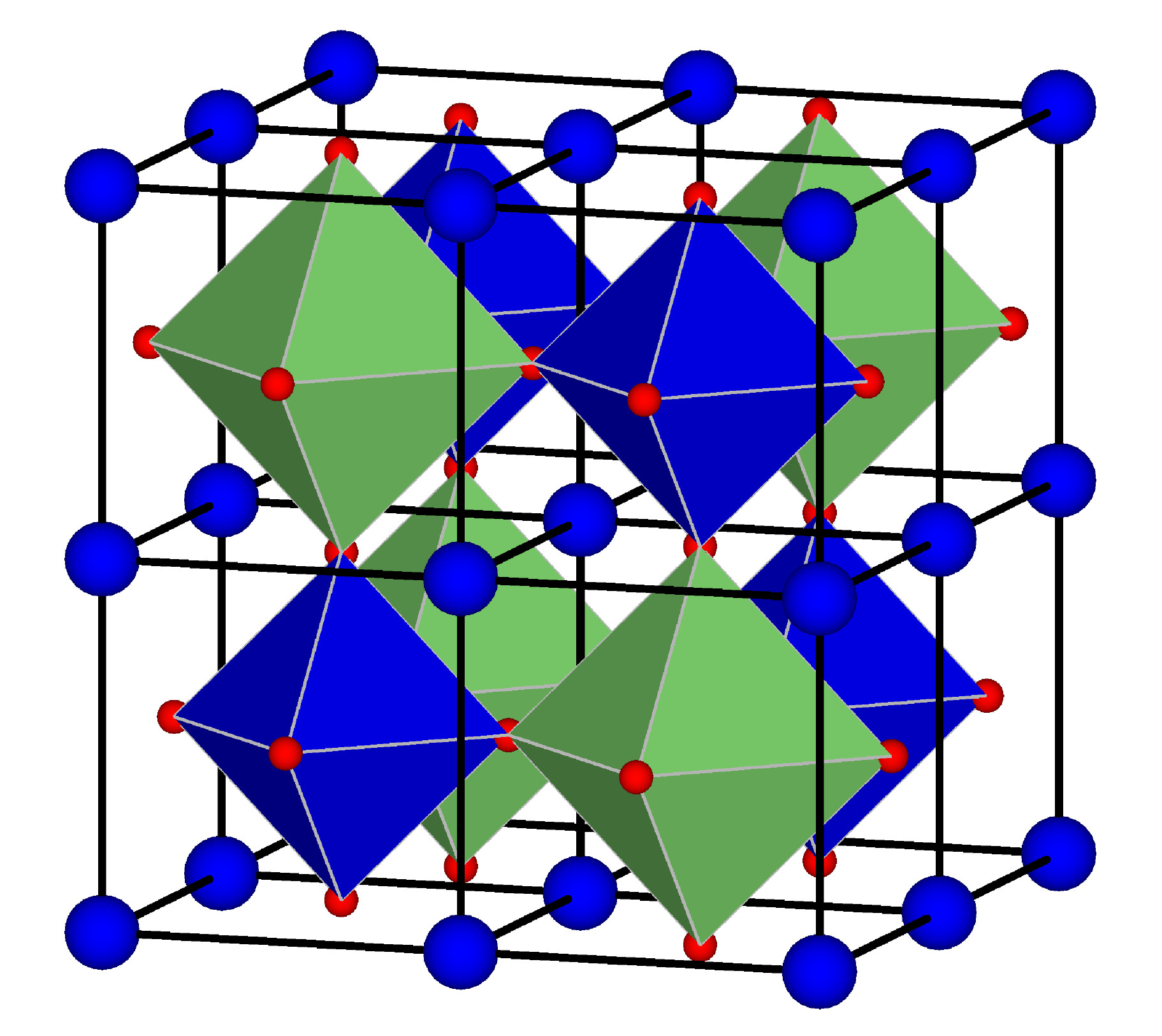}
\caption
{Sketch of the atomic positions in a cubic double-perovskite compound,
Ba$_2BB^{\prime}$O$_6$.
$B$ stands here for either Y, Na or Li (site at the center of a dark-blue octahedron);
$B^{\prime}$ is either Mo or Os (site at the center of a light-green octahedron).
O ions are shown as small red spheres while the Ba sites are the larger blue spheres.
}
\label{DP_str}
\end{figure}

Here we carry out a detailed {\it ab initio} investigation of the Mo$^{5+}$ 4$d^1$ and Os$^{7+}$
5$d^1$ relativistic electronic structure in the double-perovskite compounds Ba$_2$YMoO$_6$,
Ba$_2$LiOsO$_6$ and Ba$_2$NaOsO$_6$.
In addition to providing reliable results for the energy scale of the $d$-level splittings,
$t_{2g}$--$e_g$ and induced by SOC within the $t_{2g}^1$ manifold, we analyze the role of 
TM\,$d$\,--\,O\,$p$ orbital mixing plus the strength of electron-lattice couplings.
It is found that strong metal\,$d$\,--\,O\,$p$ hybridization generates a finite magnetic moment
even for perfectly cubic environment around the TM site, providing {\it ab initio} support to the
phenomenological covalency factor introduced in this context by Stevens
\cite{SOC_k_factor_stevens_1953}.
The TM $d^1$ magnetic moment is further enhanced by tetragonal distortions, against which the
octahedral oxygen cage is unstable.
According to our results, such Jahn-Teller (JT) effects are particularly strong for the Mo 4$d^1$
ions in Ba$_2$YMoO$_6$.
While additional investigations are needed for clarifying the role of intersite cooperative
couplings \cite{d1_soc_balents_2010,BaNaOsO_pickett_2015}, our calculations thus emphasize the high
sensitivity of the effective magnetic moments to both metal-ligand covalency effects and local 
JT physics.
The material dependence for the ratio among the strengths of the spin-orbit interaction, the JT coupling
parameter and the effective covalency factor that we compute here provide a solid basis for future
studies addressing the role of intersite interactions on the double-perovskite {\sl fcc} lattice.
\\[0.15cm]

\begin{table}[t]
\caption{
Mo$^{5+}$ 4$d$-shell splittings ($j\!=\!3/2$ to $j\!=\!1/2$ and $t_{2g}$--\,$e_{g}$) and `static' 
$g_{\parallel}$ factor in cubic Ba$_2$YMoO$_6$.
Three orbitals (4$d$ $t_{2g}$) were active in the CASSCF calculation for
$\Delta_{3/2\rightarrow 1/2}$ while
five orbitals were active in the calculations for $g_{\parallel}$ and
$\Delta_{t_{2g}\rightarrow e_{g}}$ (sans SOC for the latter).
}
\centering
\begin{tabular}{l c l c  l}
\hline\hline   \\ [-1.5ex]
4$d^1$ electronic structure
                                        &\hspace{10mm}   &CASSCF  &\hspace{10mm}  &MRCI  \\[0.5ex]
\hline \\ [-1.5ex]
$\Delta_{3/2\rightarrow1/2}$ (eV)       &\hspace{10mm}   &0.133   &\hspace{10mm}  &0.130 \\[0.5ex]
$\Delta_{t_{2g}\rightarrow e_{g}}$ (eV) &\hspace{10mm}   &4.80    &\hspace{10mm}  &4.51  \\[0.5ex]
$g_{\parallel}$                         &\hspace{10mm}   &0.18    &\hspace{10mm}  &0.20  \\[0.5ex]
\hline
\end{tabular}
\label{QC_table_Mo}
\newline
\end{table}

\noindent
{\bf {\large  RESULTS}}  \\

Quantum chemistry calculations were first performed to resolve the essential features of the
electronic structure of the cubic lattice configuration, without accounting for electron-lattice
couplings (see Methods for computational details and Fig.\,\ref{DP_str} for a sketch of a 
three-dimensional double-perovskite crystal).
Results for the splitting between the Mo$^{5+}$ $t_{2g}^1$ $j\!=\!3/2$ and $j\!=\!1/2$ spin-orbit
states, $\Delta_{\!3/2\rightarrow1/2}$, are provided in Table\,\ref{QC_table_Mo} at two levels of
approximation, i.\,e., multiconfiguration complete-active-space self-consistent-field (CASSCF) and
multireference configuration-interaction (MRCI) with single and double excitations on top of the
CASSCF wave function \cite{book_QC_00}.
Knowing the splitting $\Delta_{\!3/2\rightarrow1/2}$, the strength of the SOC parameter can be
easily derived as $\lambda\!=\!\frac{2}{3}\Delta_{3/2\rightarrow1/2}^{\mathrm{CASSCF}}$
\cite{book_abragam_bleaney}.
The resulting $\lambda$ of 89 meV is somewhat smaller than earlier estimates of 99 meV for Mo$^{5+}$
impurities in SrTiO$_3$ \cite{d1_JT_faughnan_1972}.
A most interesting finding, however, is that despite the cubic environment the quantum chemistry 
calculations yield a nonvanishing magnetic moment and a finite $g$ factor.
This obviously does not fit the nonmagnetic $j\!=\!3/2$ quartet ground state assumed to arise in
standard textbooks on crystal-field theory \cite{book_abragam_bleaney,book_mabbs_machin} from exact
cancellation between the spin and the orbital moments.

\begin{figure*} [t]
 \centering
 \subfigure{
 \hspace{-0.5cm}
\raisebox{2.2cm}{\textbf{ \large a}} \hspace{0.4cm}
\includegraphics[width=0.2\textwidth,natwidth=50,angle=0]{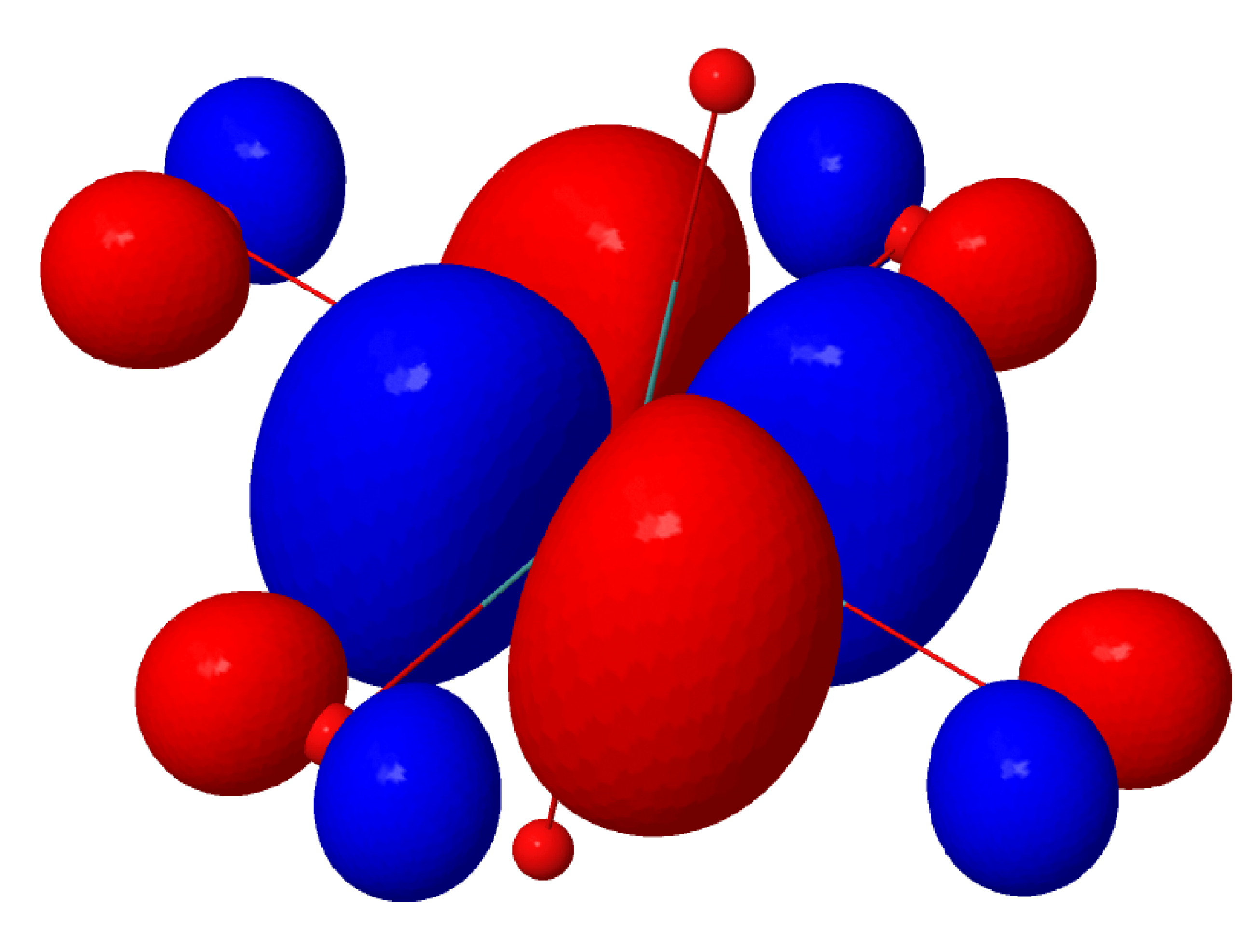}
\label{fig_dplot}}
 \quad
\subfigure{
\raisebox{2.2cm}{\textbf{ \large b}}
\includegraphics[width=0.6\textwidth,natwidth=50,angle=0]{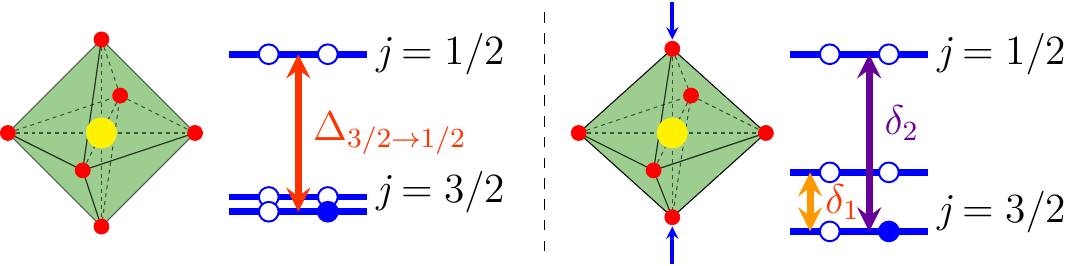}
\label{fig_E_level}} \\
\subfigure{
\raisebox{-0.25cm}{\textbf{ \large c}}
\vspace{2.8cm}
\includegraphics[width=0.25\textwidth,natwidth=50,angle=-90]{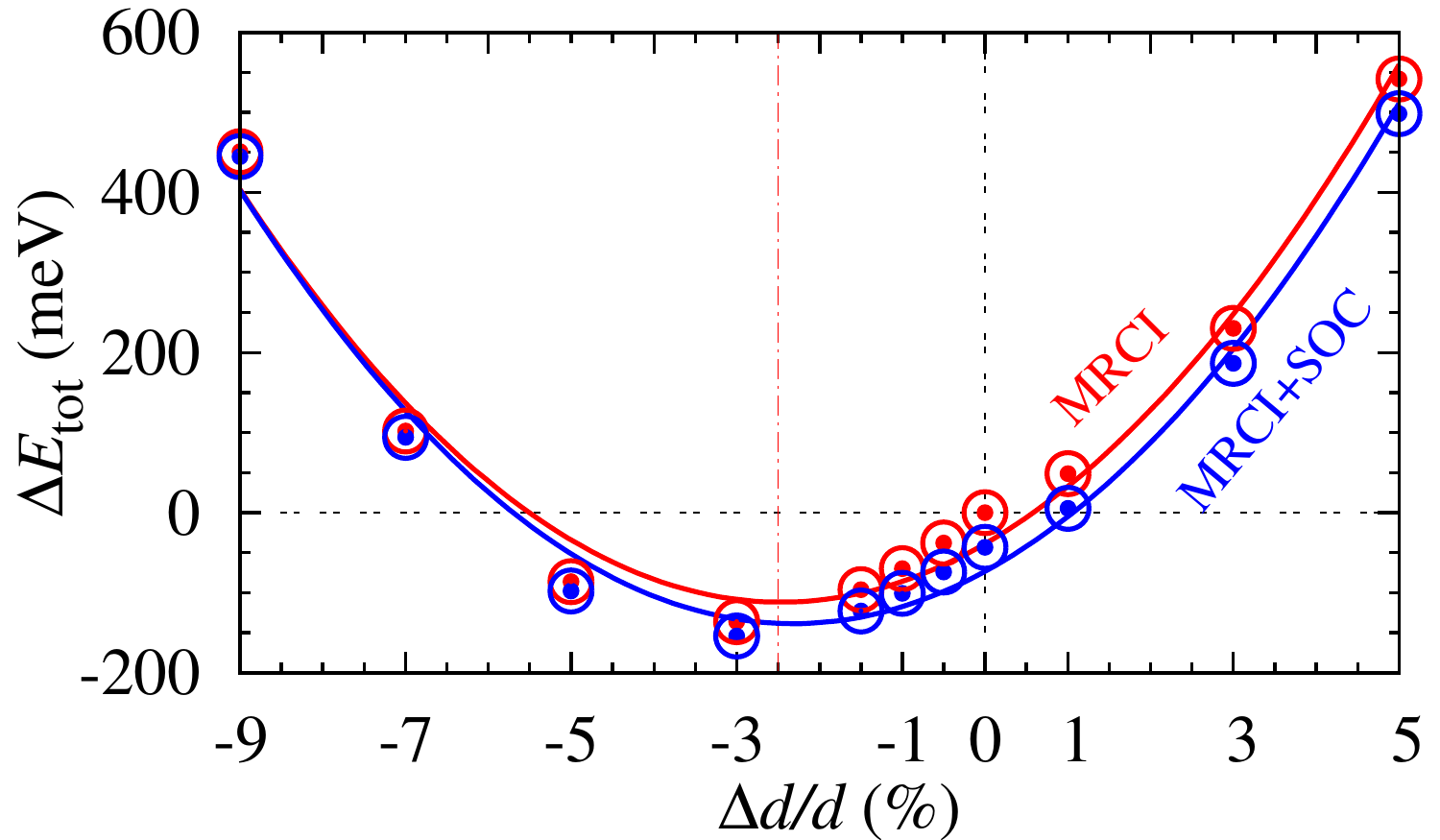}
\label{plot_tot_E}}
\quad
\subfigure{
\raisebox{-0.25cm}{\textbf{ \large d}}
\vspace{2.8cm}
\includegraphics[width=0.25\textwidth,natwidth=50,angle=-90]{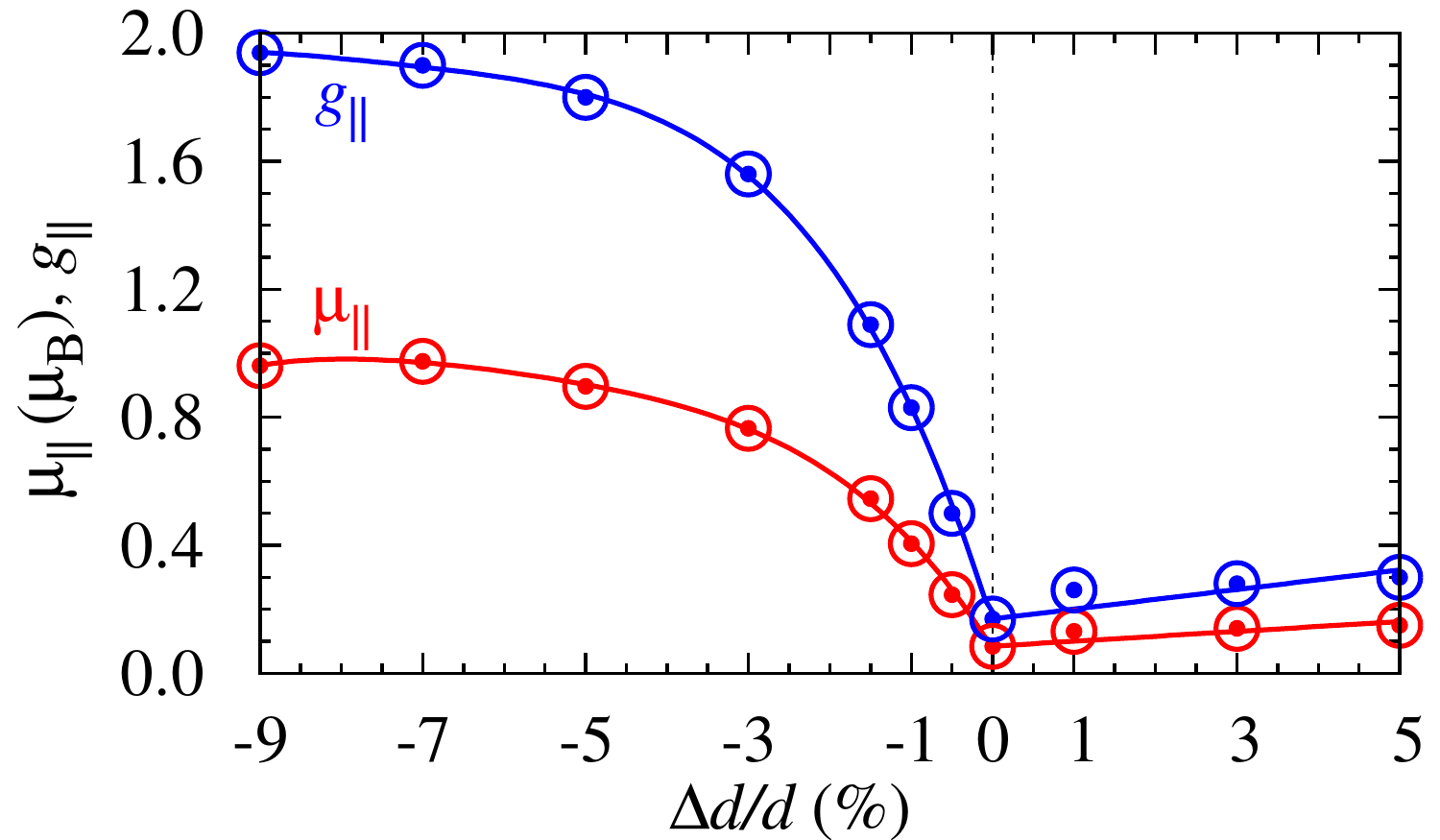}
\label{plot_M_gfactor_new}}
 \caption{
a) Mo $4d$ $t_{2g}$ charge density as obtained by CASSCF calculations.
   The tails at the nearest-neighbor O sites have substantial weight.
b) TM $t_{2g}$ splittings in cubic (left) and tetragonal (right) symmetry;
   $\delta_1\!=\!0$ and $\delta_2\!=\!\Delta_{3/2\rightarrow1/2}$ for cubic octahedra.
c) Ground-state energy as function of $z$-axis tetragonal distortion, MRCI results both with and without
   spin-orbit coupling.
d) Variation of the Mo $4d^1$ magnetic moment ($\mu_{\|}$) and $g$ factor ($g_{\|}$) with the amount
   of $z$-axis tetragonal distortion, MRCI results including spin-orbit interactions.
}
\end{figure*}

At a qualitative level, it has been argued by Stevens \cite{SOC_k_factor_stevens_1953} that finite
$g$-factor values can in fact occur for $j\!=\!3/2$ ions due to TM--O covalency on the TM\,O$_6$
octahedron.
For better insight into the nature of such effects we therefore performed a simple numerical
experiment in which the six ligands coordinating the reference Mo$^{5+}$ 4$d^1$ ion are replaced
by $-2$ point charges with no atomic basis functions.
In that additional set of computations the magnetic moment and the $g$ factor do vanish, in agreement
with the purely ionic picture of Kotani, Abragam and Bleaney \cite{d1_SOC_kotani_1949,book_abragam_bleaney}.
This shows that one tuning knob for switching magnetism on is indeed the TM\,4$d$\,--\,O\,2$p$ orbital
hybridization.
The latter is strong for high ionization states such as Mo$^{5+}$ (as the tails of the 4$d$-like
valence orbitals indicate in the case the nearest-neighbor ligands are provided with atomic basis
sets, see Fig.\,\ref{fig_dplot}, gives rise to partial quenching of the orbital moment and
makes that the exact cancellation between the spin and the orbital moments no longer holds.

We find that this effect is even stronger for the formally 7+ Os ion in Ba$_2$LiOsO$_6$ and
Ba$_2$NaOsO$_6$.
As shown in Table\,\ref{QC_table_Os}, $g$ factors as large as 0.4 are computed in this case.
The quantum chemistry results also allow us to estimate the strength of the effective Os$^{7+}$
5$d^1$ SOC constant, with
$\lambda\!=\!\frac{2}{3}\Delta_{3/2\rightarrow1/2}^{\mathrm{CASSCF}}\!=\!387$ meV, lower than
$\lambda\!=\!468$ meV in tetravalent 5$d^5$ iridates \cite{RhIr_vmk_IC_14}.

\begin{table}[b]
\caption{
Os$^{7+}$ 5$d$-shell splittings ($j\!=\!3/2$ to $j\!=\!1/2$ and $t_{2g}$--\,$e_{g}$) and `static'
$g_{\parallel}$ factors in cubic Ba$_2$LiOsO$_6$ and Ba$_2$NaOsO$_6$.
Only the 5$d$ $t_{2g}$ orbitals were active in the CASSCF calculation for
$\Delta_{3/2\rightarrow 1/2}$;
all five 5$d$ orbitals were active in the calculations for $g_{\parallel}$ and
$\Delta_{t_{2g}\rightarrow e_{g}}$.
For the latter, values including SOC are provided within parantheses.
All energies in eV.
}
\centering
\begin{tabular}{lllll}
\hline\hline   \\ [-1.5ex]
5$d^1$ electronic structure
                                             &\hspace{10mm} &CASSCF      &\hspace{10mm} &MRCI\\[0.5ex]
\hline \\[-1.5ex]

Ba$_2$LiOsO$_6$\,:\\
$\Delta_{3/2\rightarrow1/2}$                 &\hspace{10mm} &0.58        &\hspace{10mm} &0.56       \\[0.3ex]
$\Delta_{t_{2g}\rightarrow e_{g}}$           &\hspace{10mm} &6.17 (6.44) &\hspace{10mm} &5.95 (6.21)\\[0.3ex]
$g_{\parallel}$                              &\hspace{10mm} &0.39        &\hspace{10mm} &0.40       \\[0.7ex]

Ba$_2$NaOsO$_6$\,:\\
$\Delta_{3/2\rightarrow1/2}$                 &\hspace{10mm} &0.58        &\hspace{10mm} &0.57       \\[0.3ex]
$\Delta_{t_{2g}\rightarrow e_{g}}$           &\hspace{10mm} &6.41 (6.68) &\hspace{10mm} &6.19 (6.45)\\[0.3ex]
$g_{\parallel}$                              &\hspace{10mm} &0.31        &\hspace{10mm} &0.40       \\[0.3ex]

\hline
\end{tabular}
\label{QC_table_Os}
\newline
\end{table}

Since the $t_{2g}^1$ electron configuration is susceptible to JT effects, we carried out further
investigations on the stability of an ideal TM\,O$_6$ octahedron against tetragonal ($z$-axis)
distortions.
A total-energy profile for specified geometric configurations is provided in Fig.\,\ref{plot_tot_E}
for an embedded MoO$_6$ octahedron.
It is seen that the minimum corresponds to about 3\% tetragonal compression, as compared to the
cubic octahedron of the $Fm3m$ crystalline structure \cite{BaYMoO_aharen_2010}.
As expected, the magnetic moment rapidly increases in the presence of distortions, as illustrated
in Table\,\ref{lat-para} and Fig.\,\ref{plot_M_gfactor_new}\,.

Depending on further details related to the strength of the intersite couplings among `JT centers',
{\it static} deformations away from cubic symmetry may be realized in some systems, as observed 
for example in the Re$^{6+}$ 5$d^1$ double perovskite Sr$_2$MgReO$_6$ \cite{SrMgReO_wiebe_2003}
and rare-earth molibdates \cite{RE_Mo_DP_mclaughlin_2008,RE_Mo_DP_mclaughlin_2013}.
If the local JT couplings and intersite interactions are relatively weak, one may be left on the
other hand in a {\it dynamic} JT regime, as earlier pointed out for the particular $t_{2g}^1$
configuration by, e.g., Kahn and Kettle \cite{d1_JT_kahn_kettle_1975}.
The relevant vibrational modes that couple to the $^2T_{2g}$($t_{2g}^1$) electronic term are those
of $E_g$ symmetry [$(3z^2\!-\!r^2)$-\, and $(x^2\!-\!y^2)$-like].
From the quantum chemistry calculations, we find that the potential-energy well is significantly
shallower for these normal coordinates, as compared to $z$-axis-only compression.
The value we computed for the Mo$^{5+}$ ion in Ba$_2$YMoO$_6$, $\approx$40 meV, is comparable to 
the estimate made in the 1970's for Mo$^{5+}$ $t_{2g}^1$ impurity ions within the SrTiO$_3$ matrix,
$\approx$60 meV \cite{d1_JT_faughnan_1972}.

For the osmates, the depth of this potential well is much reduced, with $E_{\mathrm{JT}}$ values
in the range of 10--15 meV by spin-orbit MRCI calculations (see Table\,\ref{g_factor_table}).

The vibronic model of Kahn and Kettle \cite{d1_JT_kahn_kettle_1975} provides specific expressions
for the $g$ factors.
In particular, $g_{\parallel}$ can be parametrized as \cite{d1_JT_kahn_kettle_1975}
\begin{equation} 
g_{\parallel} = 2(1 - k_{\mathrm{cov}}k_{\mathrm{vib}}) \,,
\end{equation} 
where $k_{\mathrm{cov}}$ is Stevens' covalency factor \cite{SOC_k_factor_stevens_1953} and
\begin{equation}
k_{\mathrm{vib}} = \mathrm{exp}[-x/(1+\rho)] \,.
\end{equation}
The parameters $x$ and $\rho$ are defined as \cite{d1_JT_kahn_kettle_1975}
$x\!=\!3E_{\mathrm{JT}}/2\bar{h}\omega$ and
$\rho\!=\!3\lambda/2\bar{h}\omega$,
where $\bar{h}\omega$ is the $E_g$-mode vibrational energy, and $g_{\perp}\!=\!0$ by symmetry
\cite{book_abragam_bleaney,SOC_k_factor_stevens_1953,d1_JT_kahn_kettle_1975}.
Recent infrared transmission spectra indicate that $\bar{h}\omega\!\approx\!560$ cm$^{-1}$\,$\approx$\,70 
meV for the bond stretching phonons \cite{BaYMoO_vib_qu_2013,CaAlNbO_vib_prosandeev_2005}.
The effective parameter $k_{\mathrm{cov}}$ we can easily evaluate from the static $g_{\parallel}$
values obtained in the MRCI spin-orbit treatment (see Tables\,\ref{QC_table_Mo} and \,\ref{QC_table_Os}) if vibronic
interactions are neglected ($k_{\mathrm{vib}}\!=\!1$ for `frozen' cubic octahedra), with
$k_{\mathrm{cov}}\!\equiv\!1\!-\!g_{\parallel}^{\mathrm{MRCI}}/2$.
This yields covalency reduction factors of 0.90 for Ba$_2$YMoO$_6$ and 0.80 for the osmates.

\begin{table}[t]
\caption{
Mo$^{5+}$ $t^{1}_{2g}$ electronic structure with `static' tetragonal squeezing of the
reference MoO$_6$ octahedron.
Only the $t^{1}_{2g}$ configuration was considered in the reference CASSCF.
$\Delta_{t_{2g}}$ is $t_{2g}$ tetragonal splitting without SOC, $\delta_1$
and $\delta_2$ are excitation energies within the $t^{1}_{2g}$ manifold with
SOC accounted for ($\delta_1\!=\!0$ and $\delta_2\!=\!\Delta_{3/2\rightarrow1/2}$ for
cubic octahedra, see Fig.\,2(b)).
MRCI results, all energies in eV.
}
\centering
\begin{tabular}{lc c c c c c c c}
\hline\hline   \\ [-1.5ex]
MoO$_6$ flattening     & \hspace{5mm} &  $0.5\%$ & \hspace{6mm}  & $1.5\%$ & \hspace{6mm}  &  $3\%$  & \hspace{6mm} &  $5\%$     \\ [0.5ex]   \hline \\ [-1.5ex]
$\Delta_{t_{2g}}$      & \hspace{5mm} &  0.02    & \hspace{6mm}  &  0.07   & \hspace{6mm}  &  0.15   & \hspace{6mm} &  0.27     \\ [0.5ex]
$\delta_1$             & \hspace{5mm} &  0.02    & \hspace{6mm}  &  0.05   & \hspace{6mm}  &  0.12   & \hspace{6mm} &  0.24     \\ [0.5ex]
$\delta_2$             & \hspace{5mm} &  0.14    & \hspace{6mm}  &  0.17   & \hspace{6mm}  &  0.23   & \hspace{6mm} &  0.33     \\ [0.5ex]
$\mu_{\parallel}$ ($\mu_{\mathrm{B}}$)
                       & \hspace{5mm} &  0.24    & \hspace{6mm}  &  0.55   & \hspace{6mm}  &  0.75   & \hspace{6mm} &  0.89     \\ [0.5ex]
$g_{\parallel}$ ($k_{\mathrm{vib}}\!=\!1$)
                       & \hspace{5mm} &  0.50    & \hspace{6mm}  &  1.09   & \hspace{6mm}  &  1.55   & \hspace{6mm} &  1.80     \\ [1ex]
\hline
\end{tabular}
\label{lat-para}
\end{table}

Estimates for $g_{\parallel}$ are provided in Table\,\ref{g_factor_table}, using the Kahn-Kettle
vibronic model and the quantum chemistry results for $\lambda$, $k_{\mathrm{cov}}$ and
$E_{\mathrm{JT}}$.
It is seen that a large $\rho/x$ ratio (i.e., large $\lambda/E_{\mathrm{JT}}$) makes that 
$g_{\parallel}$ is generated mostly through covalency effects in the osmates, with minor contributions
from vibronic couplings.
On the other hand, the small $\rho/x$ ratio in Ba$_2$YMoO$_6$ gives rise to a strong enhancement
of $g_{\parallel}$ through vibronic effects, with a factor of nearly 4 between 
$(1\!-\!k_{\mathrm{cov}}k_{\mathrm{vib}})$ and $(1\!-\!k_{\mathrm{cov}})$.
This way, the interesting situation arises that the TM magnetic moment is mainly due to vibronic
effects in Ba$_2$YMoO$_6$ and predominantly to strong covalency in Ba$_2$LiOsO$_6$ and Ba$_2$NaOsO$_6$.
\\

 \

 \

 \ 

\noindent
{\bf {\large DISCUSSION}}  \\

Experimentally, the measured magnetic moments are indeed significantly smaller in Ba$_2$LiOsO$_6$
and Ba$_2$NaOsO$_6$ \cite{BaMOsO_erickson_2007,BaMOsO_steele_2011} as compared to Ba$_2$YMoO$_6$
\cite{BaYMoO_cussen_2006,BaYMoO_aharen_2010,BaYMoO_vib_qu_2013,d1_JT_faughnan_1972}.
With regard to the estimates we make here for $g_{\parallel}$, possible sources of errors concern
the accuracy of the calculated $E_{\mathrm{JT}}$ when using the experimental crystal structure as 
reference and correlation and polarisation effects beyond a single TM\,O$_6$ octahedron
\cite{incr_doll_nio_1997,incr_chan_schuetz_2014}.
The latter effects would only increase $E_{\mathrm{JT}}$.
With respect to the former aspect, it is known that by advanced quantum chemistry calculations the lattice
constants of TM oxides can be computed with deviations of less than 0.5$\%$ from the measured values
\cite{incr_doll_nio_1997}, which implies rather small corrections to $E_{\mathrm{JT}}$.
Interestingly, recent findings of additional phonon modes at low temperatures \cite{BaYMoO_vib_qu_2013}
indicate static distortions of the MoO$_6$ octahedra in Ba$_2$YMoO$_6$ and indeed a rather large
$E_{\mathrm{JT}}$.
More detailed investigations on this matter are left for future work.
Valuable experimental data that can be directly compared to our calculations would be the results of
electron spin resonance measurements of the $g$ factors.

It is also worth pointing out that using the Kahn-Kettle model even a $E_{\mathrm{JT}}$ of 75 meV,
5 to 7 times larger than the values computed by MRCI for the osmates (see Table\;\ref{g_factor_table}),
still yields a rather moderate $g_{\parallel}$ factor of 0.65 for the Os 5$d^1$ ion.
Such $g_{\parallel}$ factors of 0.4--0.6 compare quite well with the low-temperature magnetic
moment derived from magnetization and muon spin relaxation measurements on Ba$_2$NaOsO$_6$,
$\approx$0.2\,$\mu_{\mathrm{B}}$ \cite{BaMOsO_erickson_2007,BaMOsO_steele_2011}.
For the Mo 4$d^1$ ion in Ba$_2$YMoO$_6$, the computed $g_{\parallel}$ factor is much more 
sensitive to variations of $E_{\mathrm{JT}}$ --- increasing $E_{\mathrm{JT}}$ from, e.g., 40 to
200 meV enhances $g_{\parallel}$ of Eq.\,(1) from $\approx$0.6 to $\approx$1.6.

\begin{table}[b]
\caption{
TM $g_{\parallel}$ factors using the Kahn-Kettle vibronic model \cite{d1_JT_kahn_kettle_1975}
and {\it ab initio} estimates for $\lambda$, $k_{\mathrm{cov}}$ and $E_{\mathrm{JT}}$.
$x\!=\!3E_{\mathrm{JT}}/2\bar{h}\omega$, $\rho\!=\!3\lambda/2\bar{h}\omega$, $\bar{h}\omega$
is set to 70 meV \cite{BaYMoO_vib_qu_2013,CaAlNbO_vib_prosandeev_2005} and
$g_{\parallel}\!=\!2(1\!-\!k_{\mathrm{cov}}k_{\mathrm{vib}})$.
}
\begin{tabular}{l c c c c c c}
\hline\hline   \\ [-1.5ex]
                  &$E_{\mathrm{JT}}$\hspace{2mm}  &$x$\hspace{2mm}    &$\rho$\hspace{2mm}
                  &$k_{\mathrm{vib}}$\hspace{2mm} &$k_{\mathrm{cov}}$\hspace{2mm}
                                                          &$g_{\parallel}$\hspace{2mm}\\
\hline\\[-1.5ex]
Mo$^{5+}$ 4$d^1$, Ba$_2$YMoO$_6$
                  &40                &0.86\hspace{2mm}  &1.90\hspace{2mm}   &0.74\hspace{2mm}  &0.90\hspace{2mm}   &0.66\hspace{2mm} \\ [1ex] 
Os$^{7+}$ 5$d^1$, Ba$_2$LiOsO$_6$
                  &10                &0.21\hspace{2mm}  &8.29\hspace{2mm}   &0.98\hspace{2mm}  &0.80\hspace{2mm}   &0.44\hspace{2mm} \\ [1ex] 
Os$^{7+}$ 5$d^1$, Ba$_2$NaOsO$_6$
                  &15                &0.32\hspace{2mm}  &8.29\hspace{2mm}   &0.97\hspace{2mm}  &0.80\hspace{2mm}   &0.45\hspace{2mm} \\ [1ex] 
\hline
\hline
\end{tabular}
\label{g_factor_table}
\end{table}

One other remarkable prediction of Kahn and Kettle \cite{d1_JT_kahn_kettle_1975} is that the
splitting of the $j\!=\!3/2$ and $j\!=\!1/2$ states is increased through vibronic couplings,
by a factor
\begin{equation}
\gamma = 1 + x \frac{3+\rho}{3(\rho^2 - 1)} \,.
\end{equation}
This effect turns out to be small in the osmates, given the small $x$ and large $\rho$ in those
compounds.
But we compute a strong modification of the $j\!=\!3/2$ to $j\!=\!1/2$ excitation energy for
Ba$_2$YMoO$_6$, from about 0.13 eV in the absence of vibronic interactions (see Table
\ref{QC_table_Mo}) to $\approx$0.20 eV with Jahn-Teller effects included ($E_{\mathrm{JT}}\!=\!40$
meV).
Experimentally the situation can be clarified by direct resonant inelastic x-ray scattering
(RIXS) measurements on Ba$_2$YMoO$_6$.
High-resolution RIXS measurements could also address the occurrence of static distortions at low
temperatures, suggested for Ba$_2$YMoO$_6$ on the basis of extra phonon modes in the low-$T$
infrared transmission spectra \cite{BaYMoO_vib_qu_2013} and for Ba$_2$NaOsO$_6$ from the integrated
entropy through the magnetic phase transition at about 7\,K \cite{BaMOsO_erickson_2007}.
According to the MRCI data in Table\,III, a reduction by 0.5--1.5$\%$ of the interatomic distances
on one set of O--Mo--O links already gives a splitting of 20--50 meV of the low-lying spin-orbit
states.
Splittings of this size should be accessible with last-generation RIXS apparatus.

Also of interest is an experimental confirmation of the unusually large $t_{2g}$--$e_g$ gap we
predict in the double-perovskite heptavalent osmates, $\gtrsim$6 eV (see Table\,\ref{QC_table_Os}).
According to the results of additional computations we carried out, the source of this exceptional
$d$-level splitting is the stabilization of the Os $t_{2g}$ states due to the large effective 
charge (formally 7+) at the nearest-neighbor Os sites.
The latter are situated on the axes along which the lobes of the $t_{2g}$ orbitals are oriented;
in contrast, the lobes of the $e_g$ functions point towards the monovalent species (Li$^{1+}$
or Na$^{1+}$).
For example, test CASSCF calculations in which the size of the point charges placed at the 12
Os and 6 alkaline-ion nearest-neighbor sites are modified from the formal ionic values 7+ and 1+
($12\!\times\!7\!+\!6\!\times\!1\!=\!90$) to 5+ and 5+ ($12\!\times\!5\!+\!6\!\times\!5\!=\!90$)
show a reduction of about 2 eV of the $t_{2g}$--$e_g$ level splitting.
Similar effects, with relative shifts and even inversion of the $d$-electron energy levels due to
charge imbalance at nearby cation sites, were recently evidenced in Sr$_2$RhO$_4$ and Sr$_2$IrO$_4$
\cite{RhIr_vmk_IC_14,Ir214_bogdanov_15}, the rare-earth 227 iridates $R_2$Ir$_2$O$_7$ 
\cite{Ir227_hozoi_14} and Cd$_2$Os$_2$O$_7$ \cite{Os227_bogdanov_12}.
The mechanism has not been thoroughly explored so far experimentally but seems to hold much potential 
in the context of orbital engineering in TM compounds.  \\

To summarize,
it is well known that nominal orbital degeneracy gives rise in 3$d$ transition-metal oxides to
subtle couplings between the electronic and lattice degrees of freedom and very rich physics.
Here we resolve the effect of electron-lattice interactions on the magnetic properties of heavier,
4$d$ and 5$d$ transition-metal ions with a formally degenerate $t_{2g}^1$ electron configuration
in the double-perovskite materials Ba$_2$YMoO$_6$, Ba$_2$LiOsO$_6$ and Ba$_2$NaOsO$_6$.
In particular, by using advanced quantum chemistry electronic-structure calculations, we reconcile
the notion of a nonmagnetic spin-orbit-coupled $t_{2g}^1$ $j\!=\!3/2$ ground state put forward by
Kotani, Abragam, Bleaney and others \cite{d1_SOC_kotani_1949,book_abragam_bleaney,book_mabbs_machin}
with the variety of magnetic properties recently observed in 4$d^1$ and 5$d^1$ double perovskites.
Our analysis shows that the sizable magnetic moments and $g$ factors found experimentally are due 
to strong TM\,$d$\,--\,ligand\,$p$ hybridization and dynamic Jahn-Teller effects,
providing new perspectives on the interplay between metal-ligand interactions and spin-orbit
couplings in transition-metal oxides.
It also highlights the proper theoretical frame for addressing the remarkably rich magnetic
properties of $d^1$ double perovskites 
\cite{BaMOsO_erickson_2007,BaMOsO_steele_2011,BaYMoO_deVries_2010,BaYMoO_carlo_2011,IrO_balents_review_14}
in particular.
Over the last two decades, vibronic couplings have unjustifiably received low attention in the case
of these intriguing materials.
\\[0.95cm]

\noindent
{\bf {\large METHODS}} \\
 
All {\it ab initio} calculations were carried out with the quantum chemistry package {\sc molpro} \cite{molpro12}.
Crystallographic data as derived in Ref.\,\cite{BaYMoO_aharen_2010} for Ba$_2$YMoO$_6$ and in
Ref.\,\cite{Stitzer_DPs} for Ba$_2$LiOsO$_6$ and Ba$_2$NaOsO$_6$ were employed.


We used effective core potentials (ECP's), valence basis functions of triple-zeta quality and
two $f$ polarization functions for the reference Mo/Os ions \cite{basis_setY-Pd,basis_set_Os} for
which the $d$-shell excitations are explicitly computed.
All-electron triple-zeta basis sets supplemented with two $d$ polarization functions
\cite{basis_set_O} were applied for each of the six adjacent O ligands.
The eight Ba nearest neighbors were in each case modeled by Ba$^{2+}$ `total-ion' pseudopotentials
(TIP's) supplemented with a single $s$ function \cite{basis_set_Ba}.
For Ba$_2$YMoO$_6$, the six nearby Y sites were described by ECP's and valence basis functions of
double-zeta quality \cite{basis_setY-Pd}.
In Ba$_2$LiOsO$_6$ and Ba$_2$NaOsO$_6$, we employed TIP's for the six nearest Li and Na cations
and sets of one $s$ and one $p$ functions \cite{basis_setLi-Na}.
The farther solid-state surroundings enter the quantum chemistry calculations at the level of a 
Madelung ionic potential.
How the complexity and accuracy of quantum chemistry calculations for an infinite solid can be
systematically increased is addressed in, e.g., Refs.\,\cite{QC_solids_Fulde_16,book_Fulde_12,
incr_doll_nio_1997,incr_chan_schuetz_2014}.

For the CASSCF calculations of the $d$-shell splittings, we used active spaces of either three
($t_{2g}$) or five ($t_{2g}$ plus $e_{g}$) orbitals.
The CASSCF optimizations were carried out for an average of either the $^2T_{2g}$($t_{2g}^1$)
or $^2T_{2g}$($t_{2g}^1$)\,+\,$^2E_{2g}$($e_{g}^1$) eigenfunctions of the scalar relativistic
Hamiltonian.
All O $2p$ and Mo/Os $4d$/$5d$ electrons on the reference TM\,O$_6$ octahedron were correlated in
the MRCI treatment.
The latter was performed with single and double substitutions with respect to the CASSCF reference,
as described in Refs.\,\cite{Knowles92,Werner88}.
The spin-orbit treatment was carried out according to the procedure described in Ref.\,\cite{SOC_molpro}.

The $g$ factors were computed following a scheme proposed by Bolvin \cite{Bolvin_2006} and
Vancoillie \cite{Vancoillie_2007}.
For a Kramers-doublet ground state $\{\psi, \bar{\psi} \}$, the Abragam-Bleaney tensor 
\cite{book_abragam_bleaney} $\textbf{G}=gg^{T}$ can be expressed in matrix form as
\begin{equation} 
\begin{aligned}
 G_{kl} 
  & = 2 \sum_{u,v=\psi,\bar{\psi}} \left\langle u|\hat{L}_{k} + g_{e}{\hat{S}_{k}}|v \right\rangle 
                                       \left\langle v|\hat{L}_{l} + g_{e}{\hat{S}_{l}}|u \right\rangle    \\                                
  & =  \sum_{m =x,y,z} \left( \varLambda_{km} + g_{e}{\sum}_{km} \right)   
                    \left( \varLambda_{lm} + g_{e}{\sum}_{lm} \right) \,, 
\end{aligned}
\end{equation} 
where $g_e$ is the free-electron $g$ factor and
\begin{equation} 
\begin{aligned}
 {\varLambda}_{kx} & = 2Re[ \langle \bar{\psi}|\hat{L}_{k}|\psi \rangle ], \hspace{2mm}
 &&{\sum}_{kx}  = 2Re[ \langle \bar{\psi}|\hat{S}_{k}|\psi \rangle ],         \\
 {\varLambda}_{ky} & = 2Im[ \langle \bar{\psi}|\hat{L}_{k}|\psi \rangle ], \hspace{2mm}
 &&{\sum}_{ky}  = 2Im[ \langle \bar{\psi}|\hat{S}_{k}|\psi \rangle ],         \\
 {\varLambda}_{kz} & = 2[ \langle \psi|\hat{L}_{k}|\psi \rangle ], \hspace{2mm}
 &&{\sum}_{kz}  = 2[ \langle \psi|\hat{S}_{k}|\psi \rangle ].
\end{aligned}
\end{equation} 
The matrix elements of $\hat{L}$ were extracted from the {\sc molpro} outputs, while
the matrix elements of $\hat{S}$ were derived using the conventional expressions
for the generalized Pauli matrices:
\begin{equation} 
\begin{aligned}
 \left( \hat{S}_{z}\right)_{MM^{\prime}}
 & = M \delta_{MM^{\prime}} \,, \\
 \left( \hat{S}_{x}\right)_{MM^{\prime}}
 & = \frac{1}{2} \sqrt{(S+M)(S-M+1)} \delta_{M-1,M^{\prime}} \\
 & + \frac{1}{2} \sqrt{(S-M)(S+M+1)} \delta_{M+1,M^{\prime}} \,, \\
 \left( \hat{S}_{y}\right)_{MM^{\prime}}
 & =  - \frac{i}{2} \sqrt{(S+M)(S-M+1)} \delta_{M-1,M^{\prime}} \\
 & + \frac{i}{2} \sqrt{(S-M)(S+M+1)} \delta_{M+1,M^{\prime}} \,, \\   
\end{aligned}
\end{equation} 
The $g$ factors were calculated as 
the positive square roots of the three eigenvalues of \textbf{G}.   \\
\\

\noindent
{\bf {\large ACKNOWLEDGEMENTS}} \\

We thank V.\,Kataev for discussions.
Part of the computations were carried out at the High Performance Computing Center (ZIH)
of the Technical University Dresden. We acknowledge financial support from the German Research
Foundation (Deutsche Forschungsgemeinschaft, DFG)--SFB-1143 and HO-4427/2.
\\[0.15cm]

\noindent
{\bf {\large COMPETING INTERESTS}} \\

The authors declare no competing financial interests.
\\[0.15cm]

\noindent
{\bf {\large CONTRIBUTIONS}} \\

L.X. carried out the {\it ab initio} quantum chemistry calculations, with assistance from N.A.B.,
A.P., P.F. and L.H.
The mapping of the {\it ab initio} data onto the effective vibronic model was performed by
L.X., P.F. and L.H.
L.H., P.F. and J.v.d.B. designed the project.
L.X. and L.H. wrote the paper, with contributions from all other coauthors.
\\[0.15cm]

\clearpage
\noindent

\bibliographystyle{achemso}
\bibliography{biblio_arvix_Dec15}

\providecommand{\latin}[1]{#1}
\makeatletter
\providecommand{\doi}
  {\begingroup\let\do\@makeother\dospecials
  \catcode`\{=1 \catcode`\}=2\doi@aux}
\providecommand{\doi@aux}[1]{\endgroup\texttt{#1}}
\makeatother
\providecommand*\mcitethebibliography{\thebibliography}
\csname @ifundefined\endcsname{endmcitethebibliography}
  {\let\endmcitethebibliography\endthebibliography}{}
\begin{mcitethebibliography}{48}
\providecommand*\natexlab[1]{#1}
\providecommand*\mciteSetBstSublistMode[1]{}
\providecommand*\mciteSetBstMaxWidthForm[2]{}
\providecommand*\mciteBstWouldAddEndPuncttrue
  {\def\EndOfBibitem{\unskip.}}
\providecommand*\mciteBstWouldAddEndPunctfalse
  {\let\EndOfBibitem\relax}
\providecommand*\mciteSetBstMidEndSepPunct[3]{}
\providecommand*\mciteSetBstSublistLabelBeginEnd[3]{}
\providecommand*\EndOfBibitem{}
\mciteSetBstSublistMode{f}
\mciteSetBstMaxWidthForm{subitem}{(\alph{mcitesubitemcount})}
\mciteSetBstSublistLabelBeginEnd
  {\mcitemaxwidthsubitemform\space}
  {\relax}
  {\relax}

\bibitem[Kim(2008)]{IrO_mott_kim_08}
Kim,~B. J.~{\it et al}. \emph{Phys. Rev. Lett.} \textbf{2008}, \emph{101},
  076402\relax
\mciteBstWouldAddEndPuncttrue
\mciteSetBstMidEndSepPunct{\mcitedefaultmidpunct}
{\mcitedefaultendpunct}{\mcitedefaultseppunct}\relax
\EndOfBibitem
\bibitem[Witczak-Krempa \latin{et~al.}(2014)Witczak-Krempa, Chen, Kim, and
  Balents]{IrO_balents_review_14}
Witczak-Krempa,~W.; Chen,~G.; Kim,~Y.~B.; Balents,~L. \emph{Annu. Rev. Condens.
  Matter Phys.} \textbf{2014}, \emph{5}, 57--82\relax
\mciteBstWouldAddEndPuncttrue
\mciteSetBstMidEndSepPunct{\mcitedefaultmidpunct}
{\mcitedefaultendpunct}{\mcitedefaultseppunct}\relax
\EndOfBibitem
\bibitem[Jackeli and Khaliullin(2009)Jackeli, and
  Khaliullin]{Ir213_KH_jackeli_09}
Jackeli,~G.; Khaliullin,~G. \emph{Phys. Rev. Lett.} \textbf{2009}, \emph{102},
  017205\relax
\mciteBstWouldAddEndPuncttrue
\mciteSetBstMidEndSepPunct{\mcitedefaultmidpunct}
{\mcitedefaultendpunct}{\mcitedefaultseppunct}\relax
\EndOfBibitem
\bibitem[Gretarsson(2013)]{Ir213_KH_gretarsson_2013}
Gretarsson,~H.~{\it et al}. \emph{Phys. Rev. B} \textbf{2013}, \emph{87},
  220407\relax
\mciteBstWouldAddEndPuncttrue
\mciteSetBstMidEndSepPunct{\mcitedefaultmidpunct}
{\mcitedefaultendpunct}{\mcitedefaultseppunct}\relax
\EndOfBibitem
\bibitem[Chun(2015)]{Ir213_KH_BJKim_2015}
Chun,~S. H.~{\it et al}. \emph{Nat.~Phys.} \textbf{2015}, \emph{11},
  462--466\relax
\mciteBstWouldAddEndPuncttrue
\mciteSetBstMidEndSepPunct{\mcitedefaultmidpunct}
{\mcitedefaultendpunct}{\mcitedefaultseppunct}\relax
\EndOfBibitem
\bibitem[Yamaji \latin{et~al.}(2014)Yamaji, Nomura, Kurita, Arita, and
  Imada]{Ir213_yamaji_2014}
Yamaji,~Y.; Nomura,~Y.; Kurita,~M.; Arita,~R.; Imada,~M. \emph{Phys. Rev.
  Lett.} \textbf{2014}, \emph{113}, 107201\relax
\mciteBstWouldAddEndPuncttrue
\mciteSetBstMidEndSepPunct{\mcitedefaultmidpunct}
{\mcitedefaultendpunct}{\mcitedefaultseppunct}\relax
\EndOfBibitem
\bibitem[Katukuri(2014)]{Na2IrO3_vmk_14}
Katukuri,~V. M.~{\it et al}. \emph{New J. Phys.} \textbf{2014}, \emph{16},
  013056\relax
\mciteBstWouldAddEndPuncttrue
\mciteSetBstMidEndSepPunct{\mcitedefaultmidpunct}
{\mcitedefaultendpunct}{\mcitedefaultseppunct}\relax
\EndOfBibitem
\bibitem[Bogdanov(2015)]{Ir214_bogdanov_15}
Bogdanov,~N. A.~{\it et al}. \emph{Nat.~Commun.} \textbf{2015}, \emph{6},
  7306\relax
\mciteBstWouldAddEndPuncttrue
\mciteSetBstMidEndSepPunct{\mcitedefaultmidpunct}
{\mcitedefaultendpunct}{\mcitedefaultseppunct}\relax
\EndOfBibitem
\bibitem[Chaloupka \latin{et~al.}(2010)Chaloupka, Jackeli, and
  Khaliullin]{Ir213_KH_chaloupka_10}
Chaloupka,~J.; Jackeli,~G.; Khaliullin,~G. \emph{Phys. Rev. Lett.}
  \textbf{2010}, \emph{105}, 027204\relax
\mciteBstWouldAddEndPuncttrue
\mciteSetBstMidEndSepPunct{\mcitedefaultmidpunct}
{\mcitedefaultendpunct}{\mcitedefaultseppunct}\relax
\EndOfBibitem
\bibitem[Kotani(1949)]{d1_SOC_kotani_1949}
Kotani,~M. \emph{J. Phys. Soc. Jap.} \textbf{1949}, \emph{4}, 293--297\relax
\mciteBstWouldAddEndPuncttrue
\mciteSetBstMidEndSepPunct{\mcitedefaultmidpunct}
{\mcitedefaultendpunct}{\mcitedefaultseppunct}\relax
\EndOfBibitem
\bibitem[Abragam and Bleaney(1970)Abragam, and Bleaney]{book_abragam_bleaney}
Abragam,~A.; Bleaney,~B. \emph{Electron Paramagnetic Resonance of Transition
  Ions}; Clarendon Press, Oxford, 1970; pp 417--426\relax
\mciteBstWouldAddEndPuncttrue
\mciteSetBstMidEndSepPunct{\mcitedefaultmidpunct}
{\mcitedefaultendpunct}{\mcitedefaultseppunct}\relax
\EndOfBibitem
\bibitem[Mabbs and Machin(1973)Mabbs, and Machin]{book_mabbs_machin}
Mabbs,~F.~E.; Machin,~D.~J. \emph{Magnetism and Transition Metal Complexes};
  Chapman and Hall, London, 1973; pp 68--84\relax
\mciteBstWouldAddEndPuncttrue
\mciteSetBstMidEndSepPunct{\mcitedefaultmidpunct}
{\mcitedefaultendpunct}{\mcitedefaultseppunct}\relax
\EndOfBibitem
\bibitem[Cussen \latin{et~al.}(2006)Cussen, Lynham, and
  Rogers]{BaYMoO_cussen_2006}
Cussen,~E.~J.; Lynham,~D.~R.; Rogers,~J. \emph{Chem.~Mater.} \textbf{2006},
  \emph{18}, 2855--2866\relax
\mciteBstWouldAddEndPuncttrue
\mciteSetBstMidEndSepPunct{\mcitedefaultmidpunct}
{\mcitedefaultendpunct}{\mcitedefaultseppunct}\relax
\EndOfBibitem
\bibitem[Aharen(2010)]{BaYMoO_aharen_2010}
Aharen,~T.~{\it et al}. \emph{Phys. Rev. B} \textbf{2010}, \emph{81},
  224409\relax
\mciteBstWouldAddEndPuncttrue
\mciteSetBstMidEndSepPunct{\mcitedefaultmidpunct}
{\mcitedefaultendpunct}{\mcitedefaultseppunct}\relax
\EndOfBibitem
\bibitem[de~Vries \latin{et~al.}(2010)de~Vries, Mclaughlin, and
  Bos]{BaYMoO_deVries_2010}
de~Vries,~M.~A.; Mclaughlin,~A.~C.; Bos,~J.-W.~G. \emph{Phys. Rev. Lett.}
  \textbf{2010}, \emph{104}, 177202\relax
\mciteBstWouldAddEndPuncttrue
\mciteSetBstMidEndSepPunct{\mcitedefaultmidpunct}
{\mcitedefaultendpunct}{\mcitedefaultseppunct}\relax
\EndOfBibitem
\bibitem[Carlo(2011)]{BaYMoO_carlo_2011}
Carlo,~J. P.~{\it et al}. \emph{Phys. Rev. B} \textbf{2011}, \emph{84},
  100404\relax
\mciteBstWouldAddEndPuncttrue
\mciteSetBstMidEndSepPunct{\mcitedefaultmidpunct}
{\mcitedefaultendpunct}{\mcitedefaultseppunct}\relax
\EndOfBibitem
\bibitem[Qu(2013)]{BaYMoO_vib_qu_2013}
Qu,~Z.~{\it et al}. \emph{J.~Appl.~Phys.} \textbf{2013}, \emph{113},
  17E137\relax
\mciteBstWouldAddEndPuncttrue
\mciteSetBstMidEndSepPunct{\mcitedefaultmidpunct}
{\mcitedefaultendpunct}{\mcitedefaultseppunct}\relax
\EndOfBibitem
\bibitem[Stitzer \latin{et~al.}(2002)Stitzer, Smith, and zur Loye]{Stitzer_DPs}
Stitzer,~K.~E.; Smith,~M.~D.; zur Loye,~H.-C. \emph{Solid State Sci.}
  \textbf{2002}, \emph{4}, 311 -- 316\relax
\mciteBstWouldAddEndPuncttrue
\mciteSetBstMidEndSepPunct{\mcitedefaultmidpunct}
{\mcitedefaultendpunct}{\mcitedefaultseppunct}\relax
\EndOfBibitem
\bibitem[Erickson(2007)]{BaMOsO_erickson_2007}
Erickson,~A. S.~{\it et al}. \emph{Phys. Rev. Lett.} \textbf{2007}, \emph{99},
  016404\relax
\mciteBstWouldAddEndPuncttrue
\mciteSetBstMidEndSepPunct{\mcitedefaultmidpunct}
{\mcitedefaultendpunct}{\mcitedefaultseppunct}\relax
\EndOfBibitem
\bibitem[Steele(2011)]{BaMOsO_steele_2011}
Steele,~A. J.~{\it et al}. \emph{Phys. Rev. B} \textbf{2011}, \emph{84},
  144416\relax
\mciteBstWouldAddEndPuncttrue
\mciteSetBstMidEndSepPunct{\mcitedefaultmidpunct}
{\mcitedefaultendpunct}{\mcitedefaultseppunct}\relax
\EndOfBibitem
\bibitem[Stevens(1953)]{SOC_k_factor_stevens_1953}
Stevens,~K. W.~H. \emph{On the Magnetic Properties of Covalent {XY$_{6}$}
  Complexes}; 1953; Vol. 219; pp 542--555\relax
\mciteBstWouldAddEndPuncttrue
\mciteSetBstMidEndSepPunct{\mcitedefaultmidpunct}
{\mcitedefaultendpunct}{\mcitedefaultseppunct}\relax
\EndOfBibitem
\bibitem[Chen \latin{et~al.}(2010)Chen, Pereira, and
  Balents]{d1_soc_balents_2010}
Chen,~G.; Pereira,~R.; Balents,~L. \emph{Phys. Rev. B} \textbf{2010},
  \emph{82}, 174440\relax
\mciteBstWouldAddEndPuncttrue
\mciteSetBstMidEndSepPunct{\mcitedefaultmidpunct}
{\mcitedefaultendpunct}{\mcitedefaultseppunct}\relax
\EndOfBibitem
\bibitem[Gangopadhyay and Pickett(2015)Gangopadhyay, and
  Pickett]{BaNaOsO_pickett_2015}
Gangopadhyay,~S.; Pickett,~W.~E. \emph{Phys. Rev. B} \textbf{2015}, \emph{91},
  045133\relax
\mciteBstWouldAddEndPuncttrue
\mciteSetBstMidEndSepPunct{\mcitedefaultmidpunct}
{\mcitedefaultendpunct}{\mcitedefaultseppunct}\relax
\EndOfBibitem
\bibitem[Helgaker \latin{et~al.}(2000)Helgaker, J{\o}rgensen, and
  Olsen]{book_QC_00}
Helgaker,~T.; J{\o}rgensen,~P.; Olsen,~J. \emph{Molecular Electronic-Structure
  Theory}; Wiley, 2000\relax
\mciteBstWouldAddEndPuncttrue
\mciteSetBstMidEndSepPunct{\mcitedefaultmidpunct}
{\mcitedefaultendpunct}{\mcitedefaultseppunct}\relax
\EndOfBibitem
\bibitem[Faughnan(1972)]{d1_JT_faughnan_1972}
Faughnan,~B.~W. \emph{Phys. Rev. B} \textbf{1972}, \emph{5}, 4925--4931\relax
\mciteBstWouldAddEndPuncttrue
\mciteSetBstMidEndSepPunct{\mcitedefaultmidpunct}
{\mcitedefaultendpunct}{\mcitedefaultseppunct}\relax
\EndOfBibitem
\bibitem[Katukuri(2014)]{RhIr_vmk_IC_14}
Katukuri,~V. M.~{\it et al}. \emph{Inorg. Chem.} \textbf{2014}, \emph{53},
  4833--4839\relax
\mciteBstWouldAddEndPuncttrue
\mciteSetBstMidEndSepPunct{\mcitedefaultmidpunct}
{\mcitedefaultendpunct}{\mcitedefaultseppunct}\relax
\EndOfBibitem
\bibitem[Wiebe(2003)]{SrMgReO_wiebe_2003}
Wiebe,~C. R.~{\it et al}. \emph{Phys. Rev. B} \textbf{2003}, \emph{68},
  134410\relax
\mciteBstWouldAddEndPuncttrue
\mciteSetBstMidEndSepPunct{\mcitedefaultmidpunct}
{\mcitedefaultendpunct}{\mcitedefaultseppunct}\relax
\EndOfBibitem
\bibitem[Mclaughlin(2008)]{RE_Mo_DP_mclaughlin_2008}
Mclaughlin,~A.~C. \emph{Phys. Rev. B} \textbf{2008}, \emph{78}, 132404\relax
\mciteBstWouldAddEndPuncttrue
\mciteSetBstMidEndSepPunct{\mcitedefaultmidpunct}
{\mcitedefaultendpunct}{\mcitedefaultseppunct}\relax
\EndOfBibitem
\bibitem[Wallace \latin{et~al.}(2013)Wallace, Colman, and
  Mclaughlin]{RE_Mo_DP_mclaughlin_2013}
Wallace,~T.~K.; Colman,~R.~H.; Mclaughlin,~A.~C. \emph{Phys. Chem. Chem. Phys}
  \textbf{2013}, \emph{15}, 8672--8677\relax
\mciteBstWouldAddEndPuncttrue
\mciteSetBstMidEndSepPunct{\mcitedefaultmidpunct}
{\mcitedefaultendpunct}{\mcitedefaultseppunct}\relax
\EndOfBibitem
\bibitem[Kahn and Kettle(1975)Kahn, and Kettle]{d1_JT_kahn_kettle_1975}
Kahn,~O.; Kettle,~S.~F.~A. \emph{Mol. Phys.} \textbf{1975}, \emph{29},
  61--79\relax
\mciteBstWouldAddEndPuncttrue
\mciteSetBstMidEndSepPunct{\mcitedefaultmidpunct}
{\mcitedefaultendpunct}{\mcitedefaultseppunct}\relax
\EndOfBibitem
\bibitem[Prosandeev \latin{et~al.}(2005)Prosandeev, Waghmare, Levin, and
  Maslar]{CaAlNbO_vib_prosandeev_2005}
Prosandeev,~S.~A.; Waghmare,~U.; Levin,~I.; Maslar,~J. \emph{Phys.~Rev.~B}
  \textbf{2005}, \emph{71}, 214307\relax
\mciteBstWouldAddEndPuncttrue
\mciteSetBstMidEndSepPunct{\mcitedefaultmidpunct}
{\mcitedefaultendpunct}{\mcitedefaultseppunct}\relax
\EndOfBibitem
\bibitem[Doll \latin{et~al.}(1997)Doll, Dolg, Fulde, and
  Stoll]{incr_doll_nio_1997}
Doll,~K.; Dolg,~M.; Fulde,~P.; Stoll,~H. \emph{Phys. Rev. B} \textbf{1997},
  \emph{55}, 10282--10288\relax
\mciteBstWouldAddEndPuncttrue
\mciteSetBstMidEndSepPunct{\mcitedefaultmidpunct}
{\mcitedefaultendpunct}{\mcitedefaultseppunct}\relax
\EndOfBibitem
\bibitem[Yang(2014)]{incr_chan_schuetz_2014}
Yang,~J.~{\it et al}. \emph{Science} \textbf{2014}, \emph{345}, 640--643\relax
\mciteBstWouldAddEndPuncttrue
\mciteSetBstMidEndSepPunct{\mcitedefaultmidpunct}
{\mcitedefaultendpunct}{\mcitedefaultseppunct}\relax
\EndOfBibitem
\bibitem[Hozoi(2014)]{Ir227_hozoi_14}
Hozoi,~L.~{\it et al.}. \emph{Phys. Rev. B} \textbf{2014}, \emph{89},
  115111\relax
\mciteBstWouldAddEndPuncttrue
\mciteSetBstMidEndSepPunct{\mcitedefaultmidpunct}
{\mcitedefaultendpunct}{\mcitedefaultseppunct}\relax
\EndOfBibitem
\bibitem[Bogdanov(2013)]{Os227_bogdanov_12}
Bogdanov,~N. A.~{\it et al}. \emph{Phys. Rev. Lett.} \textbf{2013}, \emph{110},
  127206\relax
\mciteBstWouldAddEndPuncttrue
\mciteSetBstMidEndSepPunct{\mcitedefaultmidpunct}
{\mcitedefaultendpunct}{\mcitedefaultseppunct}\relax
\EndOfBibitem
\bibitem[Werner \latin{et~al.}(2012)Werner, Knowles, Knizia, Manby, and
  Sch\"{u}tz]{molpro12}
Werner,~H.~J.; Knowles,~P.~J.; Knizia,~G.; Manby,~F.~R.; Sch\"{u}tz,~M.
  \emph{Wiley Rev: Comp. Mol. Sci.} \textbf{2012}, \emph{2}, 242--253\relax
\mciteBstWouldAddEndPuncttrue
\mciteSetBstMidEndSepPunct{\mcitedefaultmidpunct}
{\mcitedefaultendpunct}{\mcitedefaultseppunct}\relax
\EndOfBibitem
\bibitem[Peterson \latin{et~al.}(2007)Peterson, Figgen, Dolg, and
  Stoll]{basis_setY-Pd}
Peterson,~K.~A.; Figgen,~D.; Dolg,~M.; Stoll,~H. \emph{J. Chem. Phys.}
  \textbf{2007}, \emph{126}, 124101\relax
\mciteBstWouldAddEndPuncttrue
\mciteSetBstMidEndSepPunct{\mcitedefaultmidpunct}
{\mcitedefaultendpunct}{\mcitedefaultseppunct}\relax
\EndOfBibitem
\bibitem[Figgen \latin{et~al.}(2009)Figgen, Peterson, Dolg, and
  Stoll]{basis_set_Os}
Figgen,~D.; Peterson,~K.~A.; Dolg,~M.; Stoll,~H. \emph{J. Chem. Phys.}
  \textbf{2009}, \emph{130}, 164108\relax
\mciteBstWouldAddEndPuncttrue
\mciteSetBstMidEndSepPunct{\mcitedefaultmidpunct}
{\mcitedefaultendpunct}{\mcitedefaultseppunct}\relax
\EndOfBibitem
\bibitem[Dunning(1989)]{basis_set_O}
Dunning,~T.~H. \emph{J. Chem. Phys.} \textbf{1989}, \emph{90}, 1007--1023\relax
\mciteBstWouldAddEndPuncttrue
\mciteSetBstMidEndSepPunct{\mcitedefaultmidpunct}
{\mcitedefaultendpunct}{\mcitedefaultseppunct}\relax
\EndOfBibitem
\bibitem[Lim \latin{et~al.}(2006)Lim, Stoll, and Schwerdtfeger]{basis_set_Ba}
Lim,~I.~S.; Stoll,~H.; Schwerdtfeger,~P. \emph{J. Chem. Phys.} \textbf{2006},
  \emph{124}, 034107\relax
\mciteBstWouldAddEndPuncttrue
\mciteSetBstMidEndSepPunct{\mcitedefaultmidpunct}
{\mcitedefaultendpunct}{\mcitedefaultseppunct}\relax
\EndOfBibitem
\bibitem[Fuentealba \latin{et~al.}(1982)Fuentealba, Preuss, Stoll, and von
  Szentpály]{basis_setLi-Na}
Fuentealba,~P.; Preuss,~H.; Stoll,~H.; von Szentpály,~L. \emph{Chem. Phys.
  Lett.} \textbf{1982}, \emph{89}, 418--422\relax
\mciteBstWouldAddEndPuncttrue
\mciteSetBstMidEndSepPunct{\mcitedefaultmidpunct}
{\mcitedefaultendpunct}{\mcitedefaultseppunct}\relax
\EndOfBibitem
\bibitem[Fulde(2016)]{QC_solids_Fulde_16}
Fulde,~P. \emph{Nat.~Phys.} \textbf{2016}, \emph{12}, 106--107\relax
\mciteBstWouldAddEndPuncttrue
\mciteSetBstMidEndSepPunct{\mcitedefaultmidpunct}
{\mcitedefaultendpunct}{\mcitedefaultseppunct}\relax
\EndOfBibitem
\bibitem[Fulde(2012)]{book_Fulde_12}
Fulde,~P. \emph{Correlated Electrons in Quantum Matter}; World Scientific,
  Singapore, 2012\relax
\mciteBstWouldAddEndPuncttrue
\mciteSetBstMidEndSepPunct{\mcitedefaultmidpunct}
{\mcitedefaultendpunct}{\mcitedefaultseppunct}\relax
\EndOfBibitem
\bibitem[Knowles and Werner(1992)Knowles, and Werner]{Knowles92}
Knowles,~P.~J.; Werner,~H.-J. \emph{Theor. Chim. Acta} \textbf{1992},
  \emph{84}, 95--103\relax
\mciteBstWouldAddEndPuncttrue
\mciteSetBstMidEndSepPunct{\mcitedefaultmidpunct}
{\mcitedefaultendpunct}{\mcitedefaultseppunct}\relax
\EndOfBibitem
\bibitem[Werner and Knowles(1988)Werner, and Knowles]{Werner88}
Werner,~H.-J.; Knowles,~P.~J. \emph{J. Chem. Phys.} \textbf{1988}, \emph{89},
  5803--5814\relax
\mciteBstWouldAddEndPuncttrue
\mciteSetBstMidEndSepPunct{\mcitedefaultmidpunct}
{\mcitedefaultendpunct}{\mcitedefaultseppunct}\relax
\EndOfBibitem
\bibitem[Berning \latin{et~al.}(2000)Berning, Schweizer, Werner, Knowles, and
  Palmieri]{SOC_molpro}
Berning,~A.; Schweizer,~M.; Werner,~H.-J.; Knowles,~P.~J.; Palmieri,~P.
  \emph{Mol. Phys.} \textbf{2000}, \emph{98}, 1823--1833\relax
\mciteBstWouldAddEndPuncttrue
\mciteSetBstMidEndSepPunct{\mcitedefaultmidpunct}
{\mcitedefaultendpunct}{\mcitedefaultseppunct}\relax
\EndOfBibitem
\bibitem[Bolvin(2006)]{Bolvin_2006}
Bolvin,~H. \emph{ChemPhysChem} \textbf{2006}, \emph{7}, 1575--1589\relax
\mciteBstWouldAddEndPuncttrue
\mciteSetBstMidEndSepPunct{\mcitedefaultmidpunct}
{\mcitedefaultendpunct}{\mcitedefaultseppunct}\relax
\EndOfBibitem
\bibitem[Vancoillie \latin{et~al.}(2007)Vancoillie, Malmqvist, and
  Pierloot]{Vancoillie_2007}
Vancoillie,~S.; Malmqvist,~P.; Pierloot,~K. \emph{ChemPhysChem} \textbf{2007},
  \emph{8}, 1803--1815\relax
\mciteBstWouldAddEndPuncttrue
\mciteSetBstMidEndSepPunct{\mcitedefaultmidpunct}
{\mcitedefaultendpunct}{\mcitedefaultseppunct}\relax
\EndOfBibitem
\end{mcitethebibliography}
\end{document}